# Data Management: Past, Present, and Future

Jim Gray

Microsoft Research

June 1996

# Data Management: Past, Present, and Future

Jim Gray,
Microsoft Research,
301 Howard St.
San Francisco, CA 94105,
415-778-8222
Gray@Microsoft.com.DRAFT

**Abstract:** Soon most information will be available at your fingertips, anytime, anywhere.  Rapid advances in storage, communications, and processing allow us move all information into Cyberspace.  Software to define, search, and visualize online information is also a key to creating and accessing online information.  This article traces the evolution of data management systems and outlines current trends. Data management systems began by automating traditional tasks: recording transactions in business, science, and commerce. This data consisted primarily of numbers and character strings. Today these systems provide the infrastructure for much of our society, allowing fast, reliable, secure, and automatic access to data distributed throughout the world.  Increasingly these systems automatically design and manage access to the data. The next steps are to automate access to richer forms of data: images, sound, video, maps, and other media.  A second major challenge is automatically summarizing and abstracting data in anticipation of user requests. These multi-media databases and tools to access them will be a cornerstone of our move to Cyberspace.

## 1. Introduction And Overview

Computers can now store all forms of information: records, documents, images, sound recordings, videos, scientific data, and many new data formats.  We have made great strides in capturing, storing, managing, analyzing, and visualizing this data.  These tasks are generically called data management.  This paper sketches the evolution of data management systems describing six generations of data managers shown in Figure 1. The article then outlines current trends,

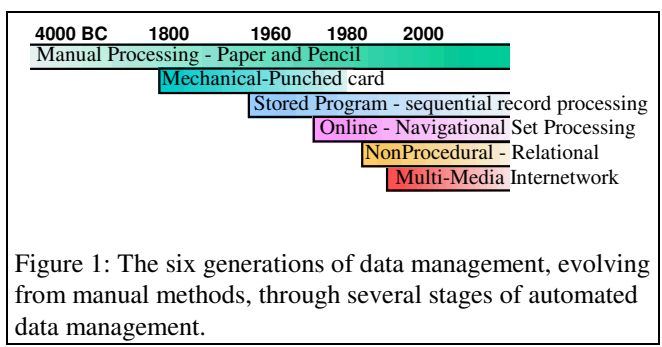

Figure 1: The six generations of data management, evolving from manual methods, through several stages of automated data management.

Data management systems typically store huge quantities of data representing the historical records of an organization.  These databases grow by accretion.  It is important that the old data and applications continue to work as new data and applications are added.  The systems are in constant change.  Indeed, most of the larger database systems in operation today were designed several decades ago and have evolved with technology.  A historical perspective helps to understand current systems.

There have been six distinct phases in data management.  Initially, data was manually processed.  The next step used punched-card equipment and electro-mechanical machines to sort and tabulate millions of records.  The third phase stored data on magnetic tape and used stored program computers to perform batch processing on sequential files.  The fourth phase introduced the concept of a database schema and online navigational access to the data.  The fifth step automated access to relational databases and added distributed and client-server processing.  We are now in the early stages of sixth generation systems that store richer data types, notably documents, images, voice, and video data.  These sixth generation systems are the storage engines for the emerging Internet and Intranets.

# 2. Historical perspective: The Six Generations of Data Management

## *2.0. Zeroth generation: Record Managers 4000BC -1900*

The first known writing describes the royal assets and taxes in Sumeria. Record keeping has a long history. The next six thousand years saw a technological evolution from clay tablets to papyrus to parchment and then to paper. There were many innovations in data representation: phonetic alphabets, novels, ledgers, libraries, paper and the printing press. These were great advances, but the information processing in this era was manual. *(Note to editor: it would be nice to have a photo of a Sumarian tablet or a Hollerith machine here. U. Penn has a good collection of photos of Sumerian tablets.)*

## *2.1. First Generation: Record Managers 1900 -1955*

The first practical automated information processing began circa 1800 with the Jacquard Loom that produced fabric from patterns represented by punched cards. Player pianos later used similar technology. In 1890, Hollerith used punched card technology to perform the US census. His system had a record for each household. Each data record was represented as binary patterns on a punched card. Machines tabulated counts for blocks, census tracts, Congressional Districts, and States. Hollerith formed a company to produce equipment that recorded data on cards, sorted, and tabulate the cards [1]. Hollerith's business eventually became International Business Machines. This small company, IBM, prospered as it supplied unit-record equipment for business and government between 1915 and 1960.

By 1955, many companies had entire floors dedicated to storing punched cards, much as the Sumerian archives had stored clay tablets. Other floors contained banks of card punches, sorters, and tabulators. These machines were programmed by rewiring control panels (patch-boards) that managed some accumulator registers, and that selectively reproduced cards onto other cards or onto paper. Large companies were processing and generating millions of records each night. This would have been impossible with manual techniques. Still, it was clearly time for a new technology to replace punched cards and electro-mechanical computers.

## *2.2. Second Generation: Programmed Unit Record Equipment 1955-1970*

Stored program electronic computers had been developed in the 1940's and early 1950's for scientific and numerical calculations. At about the same time, Univac had developed a magnetic tape that could store as much information as ten thousand cards: giving huge improvements in space, time, convenience, and reliability. The 1951 delivery of the UNIVAC1 to the Census Bureau echoed the development of punched card equipment. These new computers could process hundreds of records per second, and they could fit in a fraction of the space occupied by the unit-record equipment.

Software was a key component of this new technology. It made them relatively easy to program and use. It was much easier to sort, analyze, and process the data with languages like COBOL and RPG. Indeed, standard packages began to emerge for common business applications like general-ledger, payroll, inventory control, subscription management, banking, and document libraries.

The response to these new technologies was predictable. Large businesses recorded even more information, and demanded faster and faster equipment. As prices declined, even medium-sized businesses began to capture transactions on cards and use a computer to process the cards against a tape-based master file.

The software of the day provided a **file-oriented record processing** model. Typical programs sequentially read several input files and produced new files as output. COBOL and several other programming languages were designed to make it easy to define these record-oriented sequential tasks. Operating systems provided the file abstraction to store these records, a job control language to run the jobs, and a job scheduler to manage the workflow.

**Batch transaction processing** systems captured transactions on cards or tape and collected them in a batch for later processing. Once a day these transaction batches were sorted. The sorted transactions were merged with the much larger database (master file) stored on tape to produce a new master file. This master file also produced a report that was used as the ledger for the next day's business. Batch processing used computers very efficiently, but it had two serious shortcomings. If there was an error in a transaction, it was not detected until that evening's run against the master file, and the transaction might take several days to correct. More significantly, the business did not know the current state of the database – so transactions were not really processed until the next morning. Solving these two problems required the next evolutionary step, online systems. This step also made it much easier to write applications.

## *2.3. Third Generation: Online Network Databases 1965-1980*

Applications like stock-market trading and travel reservation need to know the current information. They could not use the day-old information provided by off-line batch transaction processing – rather they need immediate access to current data. Starting in the late 1950's, leaders in several industries began innovating with online transaction databases which interactively processed transactions against online databases. Several technologies were key to enabling online data access. The hardware to connect interactive computer terminals to a computer evolved from teletypes, to simple CRT displays, and to today's intelligent terminals based on PC technology. **Teleprocessing monitors** provided the specialized software to multiplex thousands of terminals onto the modest server computers of the day. These TP monitors collected request messages from a terminal, quickly dispatched server programs to process each message, and then dispatched the response back to the requesting terminal. **Online transaction processing** augmented the batch transaction processing that performed background reporting tasks.

Online databases stored on magnetic disks or drums provided sub-second access to any data item. These devices and data management software allowed programs to read a few records, update them, and then return the new values to the online user. Initially, the systems provided simple record lookup: either by direct lookup by record number or associative lookup by a record key.

Simple indexed-sequential record organizations soon evolved to a more powerful **set-oriented record model.** Applications often want to relate two or more records. Figure 2.a shows some record types of a simple airline reservation system and their relationships. Each city has a set of outgoing flights. Each customer has a set of trips, and each trip consists of a set of flights. In addition, each flight has a set of passengers. This information can be represented as three set-hierarchies, as shown in figure 2.b. Each of the three hierarchies answers a different question: the first is the flight schedule by city. The second hierarchy gives the customer's view of his flights. The third hierarchy tells which customers are on each flight. The travel reservation application needs all three of these data views.

The hierarchical representation of figure 2.b has a major shortcoming. Storing data redundantly is expensive, but also creates update problems: when a flight is created or is altered the flight information must be updated in all three places (all three hierarchies.) To solve these problems, the information could be represented with a **network data model** shown in figure 2.c. Figure 2.c depicts a single database where each record is stored once and is related to a set of other records via a relationship. For example, all the flights involved in a specific customer's trip are related to that trip. A program can ask the database system to enumerate those flights. New relationships among records can be created as needed. Figure 2.c is variously called a Bachman diagram or an Entity-Relationship diagram [2], [5]. The relational diagram of figure 2 (figure 2.d) is described in the next section.

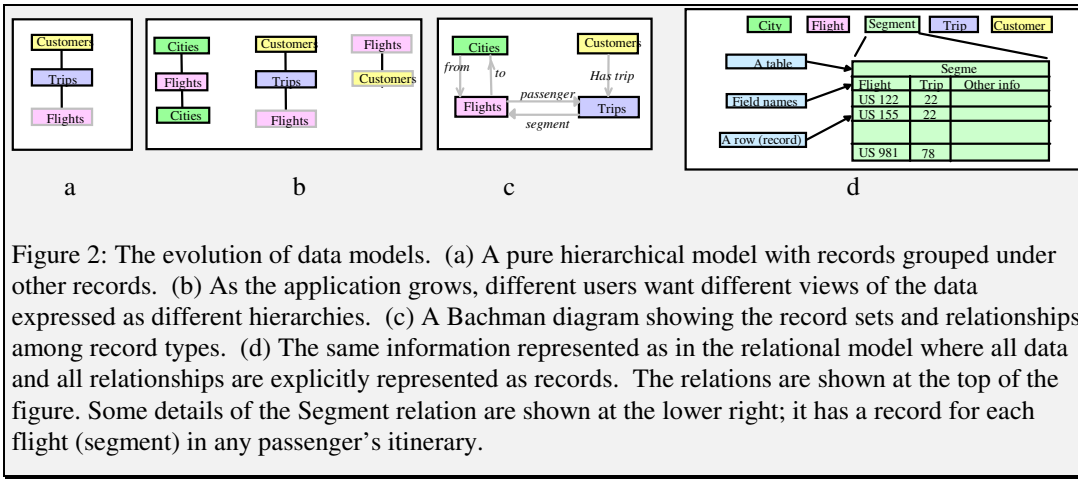

Figure 2: The evolution of data models. (a) A pure hierarchical model with records grouped under other records. (b) As the application grows, different users want different views of the data expressed as different hierarchies. (c) A Bachman diagram showing the record sets and relationships among record types. (d) The same information represented as in the relational model where all data and all relationships are explicitly represented as records. The relations are shown at the top of the figure. Some details of the Segment relation are shown at the lower right; it has a record for each flight (segment) in any passenger's itinerary.

Managing associative access and set-oriented processing was so common that the COBOL community chartered a Data Base Task Group (DBTG) to define a standard way to define and access such data. Charles Bachman had built a prototype data navigation system at GE. Bachman received the Turing award for leading the DBTG effort which defined a standard data definition and data manipulation language. In his Turing lecture he described the evolution from flat-file models to the new world where programs could navigate among records by following the relationships among the records [2]. Bachman's model is reminiscent of Vannevar Bush's Memex system [2] or the pages-and-links navigational model of today's Internet.

The COBOL database community crystallized the concept of **schemas** and data independence. They understood the need to hide the physical details of record layouts. Programs should see only the logical organization of records and relationships, so that the programs continued to work as the data layout was reorganized and evolved over time. Records, fields, and relationships not used by the program should be hidden – both for security reasons, and to insulate the program from the inevitable changes to the database design over time. These early databases supported three kinds of data schemas: (1) a **logical schema** that defines the global logical design of the database records and relationships among records, (2) a **physical schema** that describes the physical layout of the database records on storage devices and files, and the indices needed to support the logical relationships, and (3) each application was given a **sub-schema** exposing just the subset of the logical schema used by the program. The logical-physical-sub-schema mechanism provided **data independence**. Indeed, may programs written in that era are still running today using the same sub-schema the programs started with, even though the logical and physical schemas have evolved to completely new designs.

These online systems had to solve the problem of running many concurrent transactions against a database shared among many terminal users. Prior to this, the single-program-at-a-time old-master new-master approach eliminated concurrency and recovery problems. The early online systems pioneered the concept of **transactions** that lock just the records that they access. Transaction locking allows concurrent transactions to access different records. The systems also kept a log of the records that each transaction changed. If the transaction failed, the log was used to undo the effects of the transaction. The transaction log was also used for media recovery. If the system failed, the log was re-applied to an archive copy of the database to reconstruct the current database.

By 1980 the set-oriented network (and hierarchical) data models were very popular. Cullinet, a company founded by Bachman, was the largest and fastest-growing software company in the world.

## 2.4. Fourth Generation: Relational Databases and client-server computing 1980-1995

Despite the success of the network data model, many software designers felt that a navigational programming interface was too low-level. It was difficult to design and program these databases. E.F.

Codd's 1970 paper outlined the relational model [4] that seemed to provide an alternative to the low-level navigational interfaces. The idea of the **relational model** is to represent both entities and relationships in a uniform way. The relational data model has a unified language for data definition, data navigation, and data manipulation, rather than separate languages for each task. More importantly, the relational algebra deals with record sets (relations) as a group, applying operators to whole record sets and producing record sets as a result. The relational data model and operators gives much shorter and simpler programs to perform record management tasks. To give a concrete example, the airline database of the previous section would be represented by five tables as shown in Figure 2.d. Rather than implicitly storing the relationship between flights and trips, a relational system explicitly stores each flight-trip pair as a record in the database. This is the "Segment" table in Figure 2.d.

To find all segments reserved for customer Jones going to San Francisco, one would write the SQL query:
```
Select   Flight#
From     City, Flight, Segment, Trip, Customer
Where    Flight.to = "SF" AND
         Flight.flight# = Segment.flight#  AND
         Segment.trip# = trip.trip# AND
         trip.customer# = customer.customer# AND
         customer.name = "Jones"
```

The English equivalent of this SQL query is: "Find the flight numbers for flights to San Francisco which are a segment of a trip booked by any customer named "Jones." Combine the City, Flight, Segment, Tip, and Customer tables to find this flight." This program may seem complex, but it is vastly simpler than the corresponding navigational program.

Given this non-procedural query, the relational database system automatically finds the best way to match up records in the City, Flight, Segment, Trip, and Customer tables. The query does not depend on which relationships are defined. It will continue to work even after the database is logically reorganized. Consequently, it has much better data independence than a navigational query based on the network data model. In addition to improving data independence, relational programs are often five or ten times simpler than the corresponding navigational program.

Inspired by Codd's ideas, researchers in academe and industry experimented throughout the 1970's with this new approach to structuring and accessing databases promising dramatically easier data modeling and application programming. The many relational prototypes developed during this period converged on a common model and language. Work at IBM Research led by Ted Codd, Raymond Boyce, and Don Chamberlin and work at UC Berkeley led by Michael Stonebraker gave rise to a language called SQL. This language was first standardized in 1985. There have been two major additions to the standard since then [5], [6]. Virtually all database systems provide an SQL interface today. In addition, all systems provide unique extensions that go beyond the standard.

The relational model had some unexpected benefits beyond programmer productivity and ease-of-use. The relational model was well suited to client-server computing, to parallel processing, and to graphical user interfaces. **Client-server** application designs divide applications in two parts. The **client** part is responsible for capturing inputs and presenting data outputs to the user or client device. The **server** is responsible for storing the database, processing client requests against a database, and responding with a summary answer. The relational interface is especially convenient for client-server computing because it exchanges high-level requests and responses. SQL's high-level language minimizes communication between client and server. Today, many client-server tools are built around the Open Database Connectivity (ODBC) protocol that provides a standard way for clients to make high-level requests to servers. The client-server paradigm continues to evolve. As explained in the next section, there is an increasing trend to integrate procedures into database servers. In particular, procedural languages like BASIC and Java have been added to servers so that clients can invoke application procedures running at the server.

**Parallel database processing** was the second unanticipated benefit of the relational model. Relations are uniform sets of records. The relational model consists of operators closed under composition: each operator takes relations as inputs and produces a relation as a result. Consequently, relational operators naturally give pipeline parallelism by piping the output of one operator to the input of the next. It is rare to find long pipelines, but relational operators can often be partitioned so that each operator can be cloned *N* ways and each clone can work on *1/Nth* of the input relation. These ideas were pioneered by academe and by Teradata Corporation (now NCR). Today, it is routine for relational systems to provide hundred-fold speedups by using parallelism. Data mining jobs that might takes weeks or months to search multi-terabyte databases are done within hours by using parallelism. This parallelism is completely automatic. Designers just present the data to the database system, and the system partitions and indexes the data. Users present queries to the system (as ODBC requests) and the system automatically picks a parallel plan for the query and executes it.

Relational data is also well suited for **graphical user interfaces** (GUIs). It is very easy to render a relation as a set of records – relations fit a spreadsheet metaphor. Users can easily create spreadsheet-like relations and can visually manipulate them. Indeed, there are many tools that move relational data between documents, spreadsheets, and databases. Explicitly representing data, relationships, and meta-data in a uniform way makes this possible.

Relational systems combined with GUIs allow hundreds of thousands of people to pose complex database queries each day. The combinations of GUIs and relational systems has come closest to the goal of automatic-programming. GUIs allow very complex queries to be easily constructed. Given a non-procedural query, relational systems find the most efficient way to execute that query.

Continuing the historical perspective, by 1980 Oracle, Informix, and Ingress had brought relational database management systems to market. Within a few more years, IBM and Sybase had brought their products to market. By 1990, the relational systems had become more popular than the earlier set-oriented navigational systems. Meanwhile file systems, and set-oriented systems were still the workhorses of many corporations. These corporations had built huge applications over the years and could not easily change to relational systems. Rather, relational systems became the key tool for new client-server applications.

## *2.5. Fifth Generation: Multimedia Databases 1995-*

Relational systems offered huge improvements in ease-of-use, graphical interfaces, client-server applications, distributed databases, parallel data search, and data mining. Nonetheless, in about 1985, the research community began to look beyond the relational model. Traditionally, there had been a clear separation between programs and data. This worked well when the data was just numbers, characters, arrays, lists, or sets of records. As new applications appeared, the separation between programs and data became problematic. The applications needed to give the data behavior. For example, if the data was a complex object, then the methods to search, compare, and manipulate the data were peculiar to the, document, image, sound, or map datatype (see figure 3).

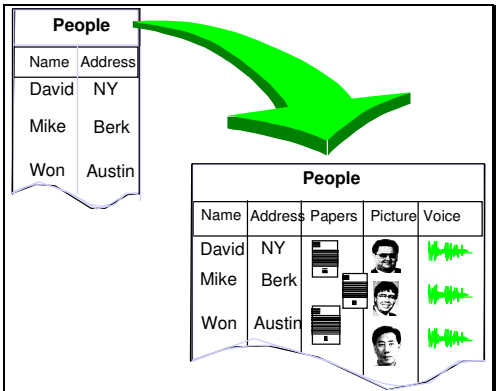

Figure 3: The transition of traditional databases storing numbers and characters into an object-relational database where each record can contain data with complex behavior. These behaviors are encapsulated in the class libraries that support the new types. In this model, the database system stores and retrieves the data and provides relationships among data items, but the class libraries provide the item behavior.

The traditional approach was to build the datatypes right into the database system. SQL added new datatypes for time, time intervals, and two-byte character strings. Each of these extensions was a significant effort. When they were done, the results were not appropriate for everyone. For example, SQL time cannot represent dates before the Christian Era and the multi-character design does not include Unicode (a universal character set for almost all languages). Users wanting Unicode or pre-Christian dates must define their own datatypes. These simple examples, and many others convinced the database community that the database system must allow domain specialists to implement the datatypes for their domains. Geographers should implement maps, text specialists should implement text indexing and retrieval, and image specialists should implement the type libraries for images. To give a specific example, a data time series is a common object type. Rather than build this object into the database system, it is recommended that the type be implemented as a class library with methods to create, update and delete a time series. Additional methods summarize trends and interpolate points in a series, and compare, combine and difference two series. Once this class library is built, it can be "plugged into" any database system. The database system will store objects of this type and will manage the data (security, concurrency, recovery, and indexing) but the datatype will manage the contents and behavior of time-series objects.

People coming from the object-oriented programming community saw the problem clearly: datatype design requires a good data model and a unification of procedures and data. Indeed, programs encapsulate the data and provide all the methods to manipulate the data. Researchers, startups, and established relational database vendors have labored long and hard since 1985 to either replace the relational model or unify the object-oriented and relational systems. Over a dozen Object-Oriented database products came to market in the late 1980's, but customers were slow to accept these systems. Meanwhile, the traditional vendors tried to extend the SQL language to embrace object oriented concepts, while preserving the benefits of the relational model.

There is still heated debate on the outcome of this evolution vs. revolution in data models. There is no debate that database systems must store and retrieve objects that are managed by class libraries. The debate revolves around the role of SQL, around the details of the object model, and around the core class libraries that the database system should support.

The rapid evolution of the Internet amplifies these debates. Internet clients and servers are being built around "applets" and "helpers" that capture, process, and render one data type or another. Users plug these applets into a browser or server. The common applets manage sound, image, text, video, spreadsheets, graphs. These applets are each class libraries for their associated types. Desktops and web browsers are ubiquitous sources and destinations for much of the data. Hence, the types and object models used on the desktop will drive the server class libraries that database systems must support.

To summarize, databases are being called upon to store more than just numbers and text strings. They are being used to store the many kinds of objects we see on the World Wide Web, and to store relationships among them. The distinction between the database and the rest of the web is being blurred. Indeed, each database vendor is promising a "universal server" that will store and analyze all forms of data (all class libraries and their objects).

Unifying procedures and data extends the traditional client-server computing model in two interesting ways: (1) active databases and (2) workflow. Active databases autonomously perform tasks when the database changes. The idea is that a user-defined trigger procedure fires when a database condition becomes true.

Using the database procedure language, database designers can define pre-conditions and triggers procedures. For example, if a re-order trigger has been defined on an inventory database, then the database will invoke a reorder procedure on an item anytime the item's inventory falls below the reorder threshold. Triggers simplify applications by moving logic from the applications to the data. The trigger mechanism is a powerful way to build active databases that are self-managing.

Workflow generalizes the typical request-response model of computing. A workflow is a script of tasks that must be executed. For example, a simple purchase agreement consists of a seven step workflow for: (1) buyer request, (2) bid, (3) agree, (4) ship, (5) invoice, (6) pay. Systems to script, execute and mange workflows are becoming common.

To close on the current status of data management technology, it makes sense to describe two large data management projects that stretch the limits of our technology today. The Earth Observation System Data / Information System (EOS/DIS) is being built by NASA and its contractors to store all the satellite data that will start arriving from the Mission to Planet Earth satellites in 1997. The database, consisting of remote sensor data, will grow by 5 terabytes a day (a terabyte is a million megabytes). By 2007, the database will have grown to 15 petabytes. This is a thousand times larger than the largest online databases today. NASA wants this database to be available to everyone, everywhere, all the time. Anyone should be able to search, analyze, and visualize the data in this database. Building EOS/DIS will require advances in data storage, data management, data search, and data visualization. Most of the data has both spatial and temporal characteristics, so the system requires substantial advances storing those data types, as well as class libraries for the various scientific data sets. For example, this application will need a library to recognize snow cover, vegetation index, clouds, and other physical features in LandSat images. This class library must easily plug into the EOS/DIS data manager.

The emerging world-wide library gives another challenging database example. Many institutional libraries are putting their holdings online. New scientific literature is being published online. Online publishing poses difficult societal issues about copyrights and intellectual property, but it also poses deep technical challenges. The size and diversity of this information are daunting. The information appears in many languages, in many data formats, and in huge volumes. Traditional or approaches to organizing this information (author, subject, title) do not exploit the power of computers to search documents by content, to link documents, and to cluster similar documents together. Information discovery, finding relevant information in the sea of text documents, maps, photographs, sounds, and videos, poses an exciting and challenging problem.

## *3. Reflections And Predictions*

Advances in computer hardware have enabled the evolution of data management from paper-based manual processing to modern information search engines. This progress in hardware is expected to continue for many more years.

Data management software has advanced in parallel to these hardware advances. The record and set-oriented systems gave way to relational systems that are now evolving to object-relational systems. These innovations give one of the best examples of research prototypes turning into products. The relational model, parallel database systems, active databases, and object-relational databases all came from the academic and industrial research labs. The development of database technology has been a textbook case of successful collaboration between academe and industry.

Inexpensive hardware and easy software have made computers accessible to almost everyone. It is now easy and inexpensive to create a web server or a database. Millions of people have done it. These users expect computers to automatically design and manage themselves. These users do not want to be computer operators. They expect to add new applications with almost no effort: a plug-and-play mentality. This view extends from simple desktop systems to very high-end servers. Users expect automated management with intuitive graphical interfaces for all administration, operations, and design tasks. Once the database is built and operational, users expect simple and powerful tools to browse, search, analyze and visualize the data. These requirements stretch the limits of what we know how to do today.

Many data management challenges remain, both technical and societal. Large online databases raise serious societal issues. Electronic data interchange and data mining software makes it relatively easy for a large organization to track all your financial transactions. By doing that, someone can build a very detailed profile of your interests, travel, and finances. Is this an invasion of your privacy? Indeed, it is possible to do this for almost everyone in the developed world. What are the implications of that? What are the privacy and security rules surrounding online medical records? Who should be allowed to see your records? How will copyrights work when anyone anywhere can access an electronic copy of a document? Cyberspace crosses national boundaries. What are the rights and responsibility of people operating in Cyberspace?

Our grandchildren will probably still be wrestling with these societal issues 50 years hence. The technical challenges are more tractable. There is broad consensus within the database community on the main challenges and a research agenda to attach those problems. Every five years, the database community does a self-assessment that outlines this agenda. The most recent self-assessment, called the Lagunita II report [8], emphasizes the following challenges:

- Defining the data models for new types (e.g., spatial, temporal, image, …) and integrating them with the traditional database systems.
- Scaling databases in size (to petabytes), space (distributed), and diversity (heterogeneous).
- Automatically discovering data trends, patterns, and anomalies (data mining, data analysis).
- Integrating (combining) data from multiple sources.
- Scripting and managing the flow of work (process) and data in human organizations.
- Automating database design and administration.

These are challenging problems. Solving them will open up new applications for computers both for organizations and for individuals. These systems will allow us to access and analyze all information from anywhere at any time. This easy access to information will transform the way we do science, the way we manage our businesses, the way we learn, and the way we play. It will both enrich and empower us and future generations.

Perhaps the most challenging problem is understanding the data. There is little question that most data will be online – both because it is inexpensive to store the data in computers and because it is convenient to store it in computers. Organizing these huge data archives so that people can easily find the information they need is the real challenge we face. Finding patterns, trends, anomalies, and relevant information from a large database is one of the most exciting new areas of data management [7]. Indeed, my hope is that computers will be able to condense and summarize information for us so that we will be spared the drudgery searching through irrelevant data for the nuggets we seek. The solution to this will require contributions from many disciplines.